\documentclass[10pt,superscriptaddress,twocolumn,amsmath,amssymb,aps,prl,showpacs]{revtex4-2}
\usepackage{mathrsfs}
\usepackage{graphicx}% Include figure files
\usepackage{dcolumn}% Align table columns on decimal point
\usepackage{bm}% bold math

\usepackage{graphicx}
\usepackage{epstopdf}
\usepackage{amsmath,amssymb,mathrsfs}
\usepackage{braket}
\usepackage{hyperref}
\hypersetup{
    colorlinks=true,
    linkcolor=blue,
    filecolor=blue,
    urlcolor=blue,
    citecolor=magenta
    }

\begin{document}

% Use the \preprint command to place your local institutional report
% number in the upper righthand corner of the title page in preprint mode.
% Multiple \preprint commands are allowed.
% Use the 'preprintnumbers' class option to override journal defaults
% to display numbers if necessary
%\preprint{}

%Title of paper
\title{Viscous Flow in a 1D Spin-Polarized Fermi Gas: the Role of Integrability on Viscosity}
%\title{The Role of the Effective Range in Resonantly Interacting Fermi Gases: \\ How Breaking Scale Symmetry Affects the Bulk Viscosity}

% repeat the \author .. \affiliation  etc. as needed
% \email, \thanks, \homepage, \altaffiliation all apply to the current
% author. Explanatory text should go in the []'s, actual e-mail
% address or url should go in the {}'s for \email and \homepage.
% Please use the appropriate macro foreach each type of information

% \affiliation command applies to all authors since the last
% \affiliation command. The \affiliation command should follow the
% other information
% \affiliation can be followed by \email, \homepage, \thanks as well.
\author{Jeff Maki}
\affiliation{Department of Physics and HKU-UCAS Joint Institute for Theoretical and Computational Physics at Hong Kong, The University of Hong Kong, Hong Kong, China}
\author{Shizhong Zhang}
\email[]{shizhong@hku.hk}
\affiliation{Department of Physics and HKU-UCAS Joint Institute for Theoretical and Computational Physics at Hong Kong, The University of Hong Kong, Hong Kong, China}

%Collaboration name if desired (requires use of superscriptaddress
%option in \documentclass). \noaffiliation is required (may also be
%used with the \author command).
%\collaboration can be followed by \email, \homepage, \thanks as well.
%\collaboration{}
%\noaffiliation

\date{\today}

\begin{abstract}
The transport properties of one-dimensional Fermi gases at low-temperatures are often described by the Luttinger liquid (LL) model. However, to study dissipation one needs to examine interactions beyond the LL model. In this work we provide a simple model which allows for a direct microscopic calculation of the bulk viscosity,  namely the one dimensional spin polarized p-wave Fermi gas. We calculate the bulk viscosity in both the high- and low-temperature limits. We find that the bulk viscosity is finite and consistent with the requirement of scale symmetry, in spite of the inherent integrability of the microscopic model. We argue how integrability does not forbid a finite bulk viscosity, and compare our work to previous kinetic theory calculations.
\end{abstract}
\maketitle

\paragraph{\it Introduction.---} Transport in integrable models has attracted much attention recently as they exhibit behavior that differs fundamentally from their non-integrable counterparts \cite{Kinoshita06,Schemmer19, Malvania21}. In integrable models, a large number of conservation laws qualitatively changes the long-wavelength dynamics, rendering the standard hydrodynamic approach inapplicable \cite{Rigol07, Langen15,Castro16,Collura18,Caux19, Bettelheim20, Doyon20}. 

%One-dimensional spin-polarized Fermi gas with a zero-range odd wave interaction furnishes one example of an integrable system. This system is constrained by a macroscopic number of conservation laws like its dual, the Lieb-Liniger model of interacting bosons \cite{Girardeau60, Lieb63, Cheon99, Valiente21}. Due to the large number of constraints, the dynamics of integrable models could be drastically different from the predictions of hydrodynamics \cite{Kinoshita06,Schemmer19, Malvania21}. In order to incorporate the large number of conservation laws in the dynamics, it is necessary to invoke generalized Gibbs ensembles and generalized hydrodynamics \cite{Rigol07,Ikeda13, Langen15,Castro16,Collura18,Caux19, Bettelheim20}.

However, the long-wavelength dynamics of the system are  still governed by a few transport coefficients, such as the conductivity, thermal conductivity, and bulk viscosity \cite{LandauFM}. Although these quantities were first derived in the hydrodynamic context, there has been a substantial amount of care in defining these transport quantities using linear response and Kubo formulae \cite{Luttinger64, Taylor10, Bradlyn12, Fujii20}. In this way  the definition of the long-wavelength transport coefficients provide meaningful and continuous results regardless of integrability. As an example, it is well known that these definitions show that the conductivity, thermal conductivity, and shear viscosity (for higher-dimensional models) are infinite when the models are integrable, and become finite with the inclusion of interactions that break the integrability \cite{Castella95, Zotos97,Rosch00,Tanaka22}.

Of particular interest for 1D integrable models is the bulk viscosity. Unlike the other transport coefficients, the bulk viscosity is special as it is constrained to be zero for scale and Galilean invariant systems \cite{Son07}, due to a hidden SO(2,1) symmetry \cite{Pitaevskii97, Werner06} that prevents entropy production for a large class of dynamics \cite{Maki18, Maki19}.  In addition, the microscopic definition of the bulk viscosity naturally implies that the bulk viscosity spectral function does not contain a ``Drude" peak for a large class of systems \cite{Bradlyn12, Fujii20}. For these reasons it is important to reconsider the role of the bulk viscosity in the dynamics of integrable systems. 

In this work, we explicitly evaluate the bulk viscosity for one integrable model, the spin-polarized Fermi gas with zero-range p-wave interactions in 1D \cite{Cheon99}. This system has been the focus of several experimental and theoretical efforts \cite{Cui16, Cui17, Sekino18, Zhang18, Hulet20, OHara20, Sekino21, Maki21} and provides an ideal platform for studying the interplay between scale symmetry, integrability and the bulk viscosity. We find that the bulk viscosity spectral function is indeed finite for finite interaction strengths at both high- and low-temperatures. Our results are consistent with scale symmetry, but inconsistent with previous kinetic theory calculations \cite{Matveev17, DeGottardi20}.

\paragraph{\it The Bulk Viscosity for the Spin-Polarized Fermi Gas.---} In this letter, we consider the following Hamiltonian for a spin-polarized Fermi gas :
\begin{align}
H &= \int dx \ \psi^{\dagger}(x)\left(-\frac{\partial_x^2}{2m} -\mu\right) \psi(x) \nonumber \\
&- \frac{g}{4m}\int dx \psi^{\dagger}(x)\overleftrightarrow{\partial_x}  \psi^{\dagger}(x) \psi(x) \overleftrightarrow{\partial_x}\psi(x)
\label{eq:H}
\end{align}
where $\psi(x)$ is the fermionic annihilation operator, $g$ is the odd-wave coupling constant, $m$ is the atomic mass, and $\mu$ is the chemical potential. $\overleftrightarrow{\partial_x}=(\overleftarrow{\partial_x}-\overrightarrow{\partial_x})/2$ is the symmetrized derivative that acts both on the immediate left and right. In principle, Eq.~(\ref{eq:H}) requires regularization, which is discussed in the supplementary materials (SM).

The (real part of the) bulk viscosity spectral function, $\zeta(\omega)$, can be defined in terms of the response of a fluid to a time-dependent strain of the systems \cite{Bradlyn12}. As shown in the SM, this reduces to:
\begin{equation}
\zeta(\omega) = \frac{Im\left[ \chi_{\Pi \Pi}(\omega)-\chi_{\Pi \Pi}(0)\right]}{\omega}
\label{eq:zeta_def}
\end{equation} 
where $\chi_{\Pi \Pi}(\omega)$ is the retarded correlation function for the trace of the stress-tensor  in frequency space:
\begin{equation}
\chi_{\Pi \Pi}(\omega) = \int_0^{\infty} e^{ i \left(\omega +i \delta\right) t} i \langle [\Pi(t),\Pi(0)]\rangle
\label{eq:chiJJ}
\end{equation}
The specific form of the trace of the stress-tensor, $\Pi(t)$, is defined below. This definition does not invoke the constitutive relations or any other hydrodynamic assumptions, only the conservation of momentum \cite{Bradlyn12}.

First let us consider the bulk viscosity at low temperatures. In this regime the dominant low-energy degrees of freedom are those near the two Fermi points, $\pm k_F$, where $k_F$ is the Fermi wavenumber which is related to the chemical potential: $\mu = k_F^2/2m$. There the dynamics is described by the Luttinger liquid (LL) model \cite{Luttinger63, Haldane81, Giamarchi}. To obtain the LL model associated with Eq.~(\ref{eq:H}), we employ the standard procedure of expanding the fermionic operators around the Fermi surface and obtaining the bosonized Hamiltonian $H =  H_{LL} + V_- + V_+$, including terms beyond the quadratic order \cite{Giamarchi}:
\begin{align}
H_{LL} &= \frac{v}{2} \int dx \left[ (\partial_x \varphi_L(x))^2 + (\partial_x \varphi_R(x))^2\right] \label{eq:HLL} \\
V_- &= \frac{\sqrt{2\pi}}{6}\tilde{\eta}_- \int dx \left[(\partial_x\varphi_L)^3(x)-(\partial_x\varphi_R)^3(x)\right] \\
V_+ &= \frac{\sqrt{2\pi}}{6}\tilde{\eta}_+ \int dx \left[(\partial_x\varphi_L)^2(x) \partial_x\varphi_R(x) - L \leftrightarrow R \right]
\label{eq:V}
\end{align}
where $H_{LL}$ is the standard LL Hamiltonian.

In Eqs.~(\ref{eq:HLL}-\ref{eq:V}), $v$ is the velocity of the chiral bosonic excitations, $\phi_r(x)$ with $r= -\bar{r} = \pm1$ for right and left movers respectively. The interactions $V_-$ and $V_+$ are the leading terms beyond the LL model allowed by parity: $\varphi_r(x) \to \varphi_{\bar{r}}(-x)$. These interactions couple like chiral modes, $V_-$, and opposite chiral modes, $V_+$  \cite{Pereira07}. We label such interactions as intra-band and inter-band scattering, respectively. The derivation of Eqs.~(\ref{eq:HLL}-\ref{eq:V}), as well as the exact expressions for $v$, $\tilde{\eta}_-$, $\tilde{\eta}_+$, and the LL parameter, $\mathcal{K}$, are given in the SM. We have also ignored two further interactions, $V_2$ and $V_2'$, as they are more irrelevant than $V_-$ and $V_+$, as shown in the SM.

In order to calculate the bulk viscosity we note that the stress-tensor can be defined through the current conservation law: $\partial_t J(x,t) + \partial_x \Pi(x,t) = 0$, where $J(x,t)$ is the particle current, and $\Pi(x,t)$ is the local stress-tensor. The trace of the stress-tensor, $\Pi(t)$, for our LL model can be obtained analytically:
\begin{align}\nonumber
\Pi(t)&=\sqrt{\frac{\mathcal{K}}{2\pi}} \frac{4\sqrt{2\pi}}{6}(3\tilde{\eta}_- + \tilde{\eta}_+) H 
+ 2 v \sqrt{\frac{\mathcal{K}}{2\pi}}\frac{\sqrt{2\pi}\tilde{\eta}_+}{6}\\
& \times \int dx \sum_r \left[\partial_x \phi_r(x,t) \partial_x \phi_{\bar{r}}(x,t) - \left(\partial_x \phi_r(x,t) \right)^2\right]
\label{eq:def_pi}
\end{align}
From Eq.~(\ref{eq:def_pi}), one can see that the intra- and inter-band couplings enter $\Pi(t)$ differently. The intra-band coupling only modifies the first term proportional to the total Hamiltonian \cite{V_minus_note}. Hence this term will not produce a finite bulk viscosity. This can be also seen by examining the density-density correlation function, as shown in Ref.~\cite{Pereira07}. There one can evaluate the density-density correlation function perturbatively with respect to $V_-$. The result exactly reproduces the non-interacting result but with a renormalized Fermi velocity, $v$, and a renormalized mass, $1/\tilde{\eta}_-$. Since the non-interacting gas has a vanishing bulk viscosity, the bulk viscosity for a Luttinger liquid with only band curvature corrections is zero. This result is intuitive since in the non-interacting limit $\tilde{\eta}_- = 1/m$, while $\tilde{\eta}_+ =-3 k_F \ell/(\pi m)$ \cite{V_minus_note}. 

The true non-trivial interaction effects are contained in $V_+$, the inter-band scattering. This interaction introduces a new term to $\Pi(t)$. When evaluating Eq.~(\ref{eq:zeta_def}), only this term contributes to the bulk viscosity as the commutator between $H$ and any operator vanishes in thermal equilibrium. One can evaluate the bulk viscosity from Eq.~(\ref{eq:def_pi}). The details are shown in the SM, here we state the result:
\begin{align}
\zeta(\omega) &= \frac{\mathcal{K}\tilde{\eta}_+^2}{72 v} \frac{\omega}{\tanh(\beta \omega/4)}
\label{eq:zeta_low}
\end{align}
% \approx \frac{m \ell^2 k_F}{8\pi^2} \frac{\omega}{\tanh\left(\frac{\omega}{2 T}\right)}.
where $\beta = 1/T$ is the inverse temperature. This is one of the main results of the Letter. 

Eq.~(\ref{eq:zeta_low}) is valid to leading order in perturbation theory with respect to $V_+$. In terms of the scattering volume $\ell$, the zero-frequency limit of the bulk viscosity at finite temperatures is: $\zeta(\omega \to 0) = 1/(2\pi)^2m k_F \ell^2 T$. On the other hand, at strictly zero temperature, the bulk viscosity depends linearly on the frequency: $\lim_{T \to 0} \zeta(\omega)=(m k_F/8\pi^2) \ell^2 \omega$, which is consistent with the vanishing of the density of states near the Fermi surface, $\rho(\omega) \propto \omega^{\frac{1}{2}(\mathcal{K} + 1/\mathcal{K})-1}$ \cite{Giamarchi}, i.e. there is a lack of flow. 

Moving towards the strong coupling limit, $V_-$ and $V_+$ are still the only allowed leading irrelevant interactions consistent with parity constraint. We also expect that both the LL parameter and the speed of sound are well behaved around resonance, namely they can be expanded in powers of $(n\ell)^{-1}$, where $n$ is the density, i.e. $\mathcal{K} = \bar{\mathcal{K}}\left(n^2/mT\right)$, $v= n/m \ \bar{v}\left(n^2/mT\right)$ and $\tilde{\eta}_+ = (n\ell)^{-1} \bar{\eta}_+\left(n^2/mT\right)$. We can combine these basic scaling relations to state that bulk viscosity must be of the form:
\begin{equation}
\zeta(\omega \to 0) = \frac{1}{n \ell^2} \bar{\zeta} \left(\frac{n^2}{mT}\right)
\label{eq:scaling}
\end{equation}
The behaviour of these scaling functions can in principle be obtained from Bethe Ansatz calculations \cite{xiwen}.

At high-temperatures, $T \gg k_F^2/2m$, one can also obtain explicit expressions for the bulk viscosity using the virial expansion in both weak and strong coupling limits. In both limits, Eq.~(\ref{eq:zeta_def}) reduces to the evaluation of the contact-contact correlation function just as in the 3D case \cite{Nishida19, Maki20b, Enss20,Hofmann20}. In the weak coupling limit, one finds
\begin{equation}
\zeta(\omega \to 0) \approx \frac{2}{\pi^{5/2}} \left( k_F \ell\right)^2 T^{1/2}, 
\label{eq:zeta_high}
\end{equation}
Combining Eq.~(\ref{eq:zeta_high}) with the result from low-temperature LL calculation for weak coupling, Eq.~(\ref{eq:zeta_low}), we see that the zero frequency bulk viscosity is proportional to $\ell^2$ and vanishes in the limit $\ell\to 0$, consistent with requirement of scale symmetry. In addition, $\zeta(\omega \to 0)$ exhibits a smooth temperature dependence since there is no finite temperature phase transition in 1D. 

%for . The bulk viscosity for weak interactions is shown in Fig.~(\ref{fig:zeta_1}) a) is shown as a function of temperature. The fact the bulk viscosity appears to be continuous as a function of temperature in the weakly interacting limit is appealing as there is no finite temperature phase transition in 1D. 

Near resonance when $k_F\ell\gg 1$, the high temperature bulk viscosity is given by
\begin{equation}
\zeta(\omega \to 0) \propto \frac{k_F^2}{\ell^2 T^{3/2}}\log\left(\frac{4}{\ell^2\omega e^{-\gamma_E}}\right).
\label{eq:zeta_high_2}
\end{equation}
Eq.~(\ref{eq:zeta_high_2})  is consistent with the general scaling form presented in Eq.~(\ref{eq:scaling}), up to a logarithmic factor. In this limit $\bar{\zeta}(x) \propto x^{3/2}$.

\begin{figure}
\includegraphics[scale=0.6]{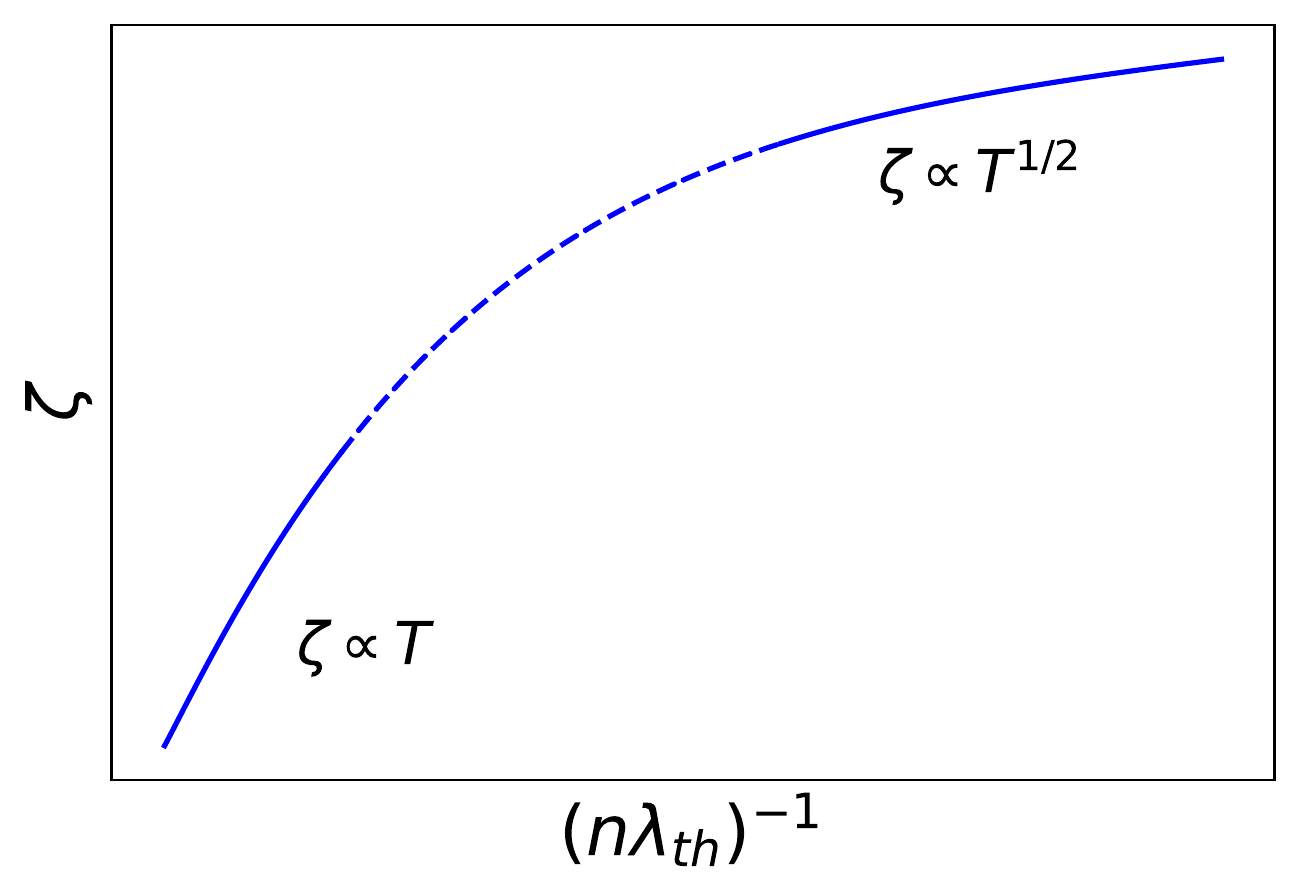}
\caption{Bulk viscosity for a weakly interacting spin-polarized 1D Fermi gas with odd-wave interactions as a function of density, $n$. $\lambda_{th} = \sqrt{2\pi / m T}$ is the thermal de Broglie wavelength. The bulk viscosity at low-temperatures is evaluated via the LL model, Eq.~(\ref{eq:zeta_low}), while the high-temperature result follows the virial expansion, Eq.~(\ref{eq:zeta_high}). In both limits the bulk viscosity is proportional to $\ell^2$ as required by conformal symmetry. Our results in the high- and low-temperature limits smoothly connect with another, consistent with the lack of a superfluid transition in 1D.}
\label{fig:zeta_1}
\end{figure}

\paragraph{\it Comparison to Traditional Kinetic Theory.---} Eqs.~(\ref{eq:zeta_low}-\ref{eq:zeta_high_2}) were calculated using the microscopic definition in Eq.~(\ref{eq:zeta_def}). As one can see these results are consistent with conformal symmetry and provide sensible results at finite and zero temperature. However, our results disagree with previous calculations using kinetic theory \cite{Matveev17, DeGottardi20}.

In the kinetic theory approach, the bulk viscosity is calculated by examining the effects of three-body interactions in a 1D spin polarized Fermi gas subjected to a slowly varying velocity gradient. According to standard arguments, the two-body scattering only leads to forward and backward scattering and can't relax the momentum, hence three-body processes are needed \cite{Khodas07,Ristivojevic16}. This argument leads to a bulk viscosity $\zeta \propto k_F^5\ell^{-2} T^{-3}$ at low-temperatures. The kinetic theory then predicts a divergent bulk viscosity in the weakly interacting limit at low-temperatures which is inconsistent with the requirement of conformal symmetry \cite{Son07}. A similar analysis at high-temperatures for zero-range p-wave interactions gives a vanishing bulk viscosity for arbitrary scattering volume because the real part of the quasiparticle self-energy leads only to the renormalization of the chemical potential and the effective mass.

It is then important to ask why the previous kinetic theory calculations are inadequate for evaluating the bulk viscosity.  Recently studies have shown that the kinetic approach is deficient in calculating the bulk viscosity, while it is accurate at describing the other transport quantities \cite{Fujii20}. The kinetic theory approach misses an important piece of the bulk-viscosity due to the correlation of pairs, as indicated explicitly in Eq.~(\ref{eq:zeta_def}). Such correlation effects do not occur at the single-particle level, and is absent in the standard kinetic theory calculation \cite{Enss_Private}. 

\paragraph{A Modified Kinetic Theory Approach.---} From our response function calculations, we know the bulk viscosity arises solely from the two-body interactions, $V_+$, which has been previously ignored \cite{Matveev17}. In principle, one may ask whether a kinetic theory calculation including $V_+$ can produce qualitatively correct results. Consider the following Hamiltonian of fermionic quasiparticles from the LL model:
\begin{align}
H_0 &= \sum_{r=\pm1}\int dx :\psi^{\dagger}_r(x)\left(-i r v \partial_x\right) \psi_r(x)\nonumber\\
&+\frac{\tilde{\eta}_-}{2} \partial_x \psi^{\dagger}(x) \partial_x \psi(x) \colon \label{eq:H0_kt}\\
V_+  &= \tilde{\eta}_+\int dx \sum_r \colon i r \psi^{\dagger}_r(x)\overleftrightarrow{\partial_x} \psi_r(x) \psi_{\bar{r}}^{\dagger}(x) \psi_{\bar{r}}(x) \colon
\label{eq:boltz_int}
\end{align}
where $\colon \ \colon$ represents normal ordering. In defining Eq.~(\ref{eq:H0_kt}) we note the first line comes from the LL Hamiltonian and defines the renormalized velocity, $v$. The second term is due to $V_-$ and describes the quadratic dispersion of the quasiparticles with an effective mass $1/\tilde{\eta_-}$. We have again neglected any term that is more irrelevant than $V_+$. 

We then follow the calculation in Ref. \cite{Matveev17} and subject the fermionic quasi-particles to a slowly varying velocity field $u(x)$. The presence of the external velocity gradient will induce non-equilibrium dynamics and dissipation. For small velocity gradients, the power dissipated can be written as: $W = \zeta L \left(\partial_x u\right)^2$. Since the power dissipated is related to entropy production, $W = T\dot{S}$, one can relate the bulk viscosity in terms of the quasiparticle distribution via:
\begin{equation}
\zeta (\partial_x u)^2= \frac{T}{L} \sum_p \frac{(\dot{n}_p^0)^2 }{n_p^0(1-n_p^0)} \tau
\label{eq:zeta_kt}
\end{equation}
where $n_p^0(t)$ is the equilibrium quasiparticle distribution function $n_p^0 = \left[\exp(\beta (\epsilon_p - \mu))+1\right]^{-1}$,  $\epsilon_p$ is the fermionic quasiparticle dispersion, $\Delta\mu = \mu(T) - \mu(0)$ is the residual chemical potential at finite $T$, and $\dot{n}_p^0(t)$ is formally of order $\partial_x u$. In defining Eq.~(\ref{eq:zeta_kt}) we have used the relaxation time approximation, and $\tau$ is the scattering rate of the quasiparticle due to $V_+$.

Now, the crucial question is to determine the value of $\tau$. Traditionally, one associates $\tau$ with three-body interactions \cite{Matveev17}. This leads to results inconsistent with our microscopic calculation. To understand how kinetic theory could produce a finite bulk viscosity, we instead associate $\tau$ with the two-particle scattering time produced by $V_+$. As shown in the SM, the scattering rate is:
\begin{equation}
\frac{1}{\tau} \approx A \frac{\tilde{\eta}_+^2T^3}{v^4}
\label{eq:tau}
\end{equation}
where $A$ is some constant and we have ignored the $p$ dependence of the inter-band scattering rate.

Given $\tau$, we turn next to the time derivative of the quasiparticle distribution function, $\dot{n}_p^0(t)$. This can be determined by noting that the quasiparticle scattering conserves number and energy. The result is:
\begin{equation}
\dot{n}_p^0 = n_p^0(1-n_p^0) \frac{3v^2 p^2 - \pi^2 T^2}{6 \tilde{\eta}_-^{-1}v^2 T} \left( n \partial_n \ln(\nu)\right) \partial_x u 
\label{eq:npdot}
\end{equation}
and $\nu = \pi n  \tilde{\eta}_- /v$. Substituting Eqs.~(\ref{eq:tau}-\ref{eq:npdot}) into Eq.~(\ref{eq:zeta_kt}) and expanding to leading order in $k_F \ell$, we obtain the final expression for the bulk viscosity in the weak coupling limit:
\begin{equation}
\zeta \approx C m k_F \ell^2 T + O(T^2)
\label{eq:zeta_kt_final}
\end{equation}
for some constant $C$ constant. Eq.~(\ref{eq:zeta_kt_final}) is consistent with the LL result, Eq.~(\ref{eq:zeta_low}). 

This result is encouraging because it is consistent with scale symmetry. However, we have assumed that the two-body scattering can lead to a relaxation of the quasiparticle distribution function. It is still unknown how these results could arise from a proper kinetic theory calculation, especially since the bulk viscosity is determined by multi-particle correlations. On the other hand, it is also important to understand within the response function formalism, how three-body interactions can affect the bulk viscosity. 

\paragraph{\it Discussions.---} We have shown in this Letter that there is indeed a bulk viscosity for a Galilean invariant integrable model, the spin polarized Fermi gas in 1D. This is interesting as naively one associates integrability with divergent transport physics. A natural question is to ask whether the dual of this model, the Lieb Liniger model of interacting bosons \cite{Girardeau60, Cheon99, Lieb63, Valiente21}, also has a finite bulk viscosity. 

From the virial expansion we can establish explicit duality relations between the fermionic bulk viscosity $\zeta_F(\ell)$ and the bosonic counterpart $\zeta_B(a)$ at high-temperatures (see SM):
\begin{equation}
\zeta_F(\ell) = \zeta_B \left(\frac{1}{a}\right).
\label{eq:duality}
\end{equation}
For these two models, the duality is of the strong-to-weak variety, mapping the strongly interacting bulk viscosity of one gas to the weakly interacting limit of the other. We do not expect the temperature to affect the duality relation. As a result, we can deduce that the bulk viscosity of the Lieb-Liniger model near resonance and at low-temperatures is given by Eq.~(\ref{eq:zeta_low}) with the replacement of $\ell$ with $a^{-1}$ via Eq.~(\ref{eq:duality}).  Similarly, the bulk viscosity for the spin-polarized Fermi gas near resonance and at low-temperatures can be determined by evaluating the bulk viscosity for a weakly interacting Bose gas in 1D. This gives a bulk viscosity  consistent with Eq.~(\ref{eq:scaling}) where $\bar{\zeta}(x) = x^4$, i.e. the bulk viscosity near resonance diverges as $T^{-4}$ \cite{Tanaka22b}.

%These results have vast consequences on the transport of one-dimensional systems. Often these integrable systems are described using dissipationless generalized hydrodynamics \cite{Caux19, Bettelheim20, Doyon20,Schemmer19}. The possibility of a finite bulk viscosity can be used alongside generalized hydrodynamics to further simulate the dynamics of integrable quantum gases in 1D. 

\paragraph{Acknowledgements}-- The authors would like to thank Xiwen Guan, Ian Affleck, Wade DeGottardi, Manuel Valiente, and Tilman Enss for useful discussions. This work is supported by HK GRF 17304820, 17304719 and CRF C6009-20G and C7012-21G. 

\paragraph{\it Note Added:} During the preparation of this manuscript, we learned of another work in preparation on this subject \cite{Tanaka22b}. There the authors calculate the bulk viscosity spectral function for the spin-polarized Fermi gas and interacting Bose gas. Their results are consistent with ours.

\newpage
\section{Supplementary Materials for Viscous Flow in a 1D Spin-Polarized Fermi Gas: the Role of Integrability on Viscosity}

The goal of this supplementary materials is to provide a detailed analysis on the bulk viscosity of a zero-ranged p-wave Fermi gas confined to one spatial dimension (1D). We begin by examining the two-body scattering properties of this system.

\section{Two-Body Scattering of the Spin-Polarized Fermi Gas}

The Hamiltonian for this system is given by:
\begin{align}
H &= \int dx \ \psi^{\dagger}(x)\left(-\frac{\partial_x^2}{2m} -\mu\right) \psi(x) \nonumber \\
&- \int dx \ \frac{g}{4m} \psi^{\dagger}(x) \overleftrightarrow{\partial_x} \psi^{\dagger}(x) \psi(x) \overleftrightarrow{\partial_x}\psi(x)
\label{eq:H_xspace}
\end{align}
where $\psi^{(\dagger)}(x)$ is the fermionic annihilation (creation) operator, $m$ is the atomic mass, $\mu$ is the chemical potential, $\overleftrightarrow{\partial_x}=(\overleftarrow{\partial_x}-\overrightarrow{\partial_x})/2$ is the symmetrized derivative that acts both on the immediate left and right, and $g$ is the bare p-wave coupling constant that depends on the ultraviolet cut-off for the theory, $\Lambda$. In this note we have set $\hbar$ to unity.

In order to regularize the theory, we solve for the two-body T-matrix. Consider two fermions with momenta $Q/2 \pm k$ and total energy $E=Q^2/4m + k^2/m$ scattering into a state of two fermions with momenta $Q/2 \pm k'$. The corresponding matrix element for the two-body T-matrix is:
\begin{align}
\langle \frac{Q}{2} \pm k | & T | \frac{Q}{2} \pm k' \rangle = k k' \tilde{T}(E) \nonumber \\
 \frac{1}{\tilde{T}(E)} &= \frac{1}{g} - \frac{1}{2}\int_{-\Lambda}^{\Lambda} \frac{dl}{2\pi} \frac{l^2}{E-\frac{Q^2}{4m}-\frac{l^2}{m}+i \delta} \nonumber \\
&= \frac{1}{g} +  \frac{m \Lambda}{2\pi} - \frac{m}{4}\sqrt{-E+Q^2/4 - i \delta}
\label{eq:T-matrix}
\end{align}
Given Eq.~(\ref{eq:T-matrix}) one can show that the scattered relative wavefunction is given by:
\begin{align}
\psi(x) &= e^{i k x} + f_k \frac{x}{|x|}e^{i k |x|} \nonumber\\
f_k& = -i k\frac{m\tilde{T}(k^2/m)}{4} 
\label{eq:scattered_wf}
\end{align}
where $f_k$ is the scattering amplitude.  According to the effective range expansion, the scattering amplitude can be written as:
\begin{equation}
f_k = i k \left[-\frac{1}{\ell}  - i k +O(k^2) \right]^{-1} 
\label{eq:effective_range}
\end{equation}
where $\ell$ is the one-dimensional p-wave scattering volume. Comparing Eq.~(\ref{eq:T-matrix}) to Eq.~(\ref{eq:effective_range}), one can identify:
\begin{equation}
\frac{m}{4\ell} = \frac{1}{g} + \frac{m \Lambda}{2\pi}.
\end{equation}
For most of this work we will focus on the weakly interacting gas. In this limit, $\ell k \ll 1$, the T-matrix can be approximated by the bare coupling constant $g$
\begin{equation}
g \approx \frac{4\ell}{m}.
\end{equation}

\section{Definition of the Bulk Viscosity}
\label{sec:def_bulk}
There are numerous studies discussing the definition of the bulk viscosity spectral function $\zeta(\omega)$ using linear response. Although several studies obtain microscopic expressions for the bulk viscosity  by invoking hydrodynamics, these expressions are actually quite general and are independent of the applicability of hydrodynamics. Here we report an equivalent definition of the bulk viscosity  \cite{Bradlyn12, Fujii20}:
\begin{align}
\zeta(\omega) &=  \frac{\chi_{\Pi,\Pi}(\omega)}{i\omega} \nonumber \\
&- \frac{1}{i \omega}\left(\left\langle \left. \frac{d\Pi_b}{d b}\right|_{b=0} \right \rangle  - \left.  \frac{d}{d b} \langle\Pi \rangle_b \right|_{b=0}\right)
\label{eq:zeta_def_app}
\end{align}
In Eq.~(\ref{eq:zeta_def_app}), the first term is the retarded stress-tensor correlation function:
\begin{equation}
\chi_{\Pi, \Pi}(\omega) =  \int_{-\infty}^{\infty} dt e^{i\left( \omega+i \delta\right) t} i \theta(t) \langle \left[\Pi(t), \Pi(0)\right]\rangle
\label{eq:PiPi_retarded}
\end{equation} 
where $\Pi$ is the stress-tensor defined below in Eq.~(\ref{def:pi}). The second term is the so called "Drude peak" or contact term. For this term we define the dilated stress-tensor, $\Pi_b$ as:
\begin{equation}
\Pi_b  =e^{i D b} \Pi e^{-i D b}
\end{equation}
where $D$ being the generator of scale transformations. The Drude peak in Eq.~(\ref{eq:zeta_def_app}) has a natural interpretation as the difference between how the stress-tensor operator behaves under a scale transformation and how the expectation value of the stress-tensor changes under a scale transformation. To make a clear connection with previous formulations of the problem, we note that $\langle \Pi\rangle = P L$, where $P$ is the pressure and $L$ is the volume. 

In order to examine the Drude peak, it is best to invoke the original microscopic model for the spin-polarized Fermi gas Eq.~(\ref{eq:H_xspace}). The stress-tensor for this model is  defined as:
\begin{equation}
\Pi = 2 H +\frac{1}{\ell} C_{\ell} 
\label{def:pi}
\end{equation}
where we also define the thermodynamic contact \cite{Maki21}:
\begin{align}
C_{\ell} &= - \frac{\partial H}{\partial \ell^{-1}} \nonumber \\
&= -\int dx \frac{g^2}{4} \psi^{\dagger}(x) \overleftrightarrow{\partial_x} \psi^{\dagger}(x) \psi(x) \overleftrightarrow{\partial_x}\psi(x)
\label{def:contact}
\end{align}
Under a scale transformation the field operator transforms as:
\begin{equation}
\psi_b(x) = e^{-b}\psi\left(x e^{-b}\right).
\end{equation}
From the transformation, the stress-tensor changes according to:
\begin{equation}
\Pi_b = 2 e^{-2b} H(\ell e^{-b}) +\frac{e^{-b}}{\ell} C_{\ell}(\ell e^{-b})
\end{equation}
Note that we have explicitly included the dependence of $\ell$ in the Hamiltonian and the contact to emphasize how these operators change under a scale transformation. It readily follows that:
\begin{align}
\left\langle \left. \frac{d\Pi_b}{d b}\right|_{b=0} \right \rangle = -2 P L - \frac{1}{\ell} \langle C_{\ell} \rangle 
\label{eq:pi_scale}
\end{align}
The second term in the Drude peak can be evaluated by noting: $\langle \Pi \rangle = P L$. From dimensional analysis \cite{Fujii20}:
\begin{equation} 
PL = \frac{1}{L^2} \bar{p}\left(\frac{L}{\ell}, S,N\right)
\end{equation}
for some dimensionless function $\bar{p}$. Under a scale transformation the length of the system changes as $L \to L e^{b}$, which corresponds to:
\begin{equation}
\left.  \frac{d}{d b} \langle\Pi_b \rangle \right|_{b=0} = -2 P L - \frac{1}{\ell} \langle C_{\ell} \rangle - \frac{1}{\ell^2}\frac{\partial}{\partial \ell^{-1}} \langle C_{\ell} \rangle
\label{eq:p_scale}
\end{equation}
The combination of Eqs.~(\ref{eq:pi_scale}) and $(\ref{eq:p_scale})$ with Eq.~(\ref{eq:zeta_def_app}) show that the Drude peak is related to the change in the contact with respect to the scattering volume. This quantity is naturally related to the stress-tensor correlation function and linear response Theory. This is evident when one considers how the contact changes under an infinitesimal time independent change in the scattering volume, $\ell \to \ell + \delta \ell$:
\begin{equation}
-\frac{\partial \langle C_{\ell} \rangle}{\partial \ell} \approx \chi_{\Pi \Pi}(0)
\end{equation}
This leads us to the final result that the bulk viscosity spectral function is given by:
\begin{equation}
\zeta(\omega) = \frac{\chi_{\Pi\Pi}(\omega)-\chi_{\Pi\Pi}(0)}{i\omega}
\label{eq:zeta_def_final}
\end{equation}

\section{Linearization and Bosonization of the Spin-Polarized Fermi Gas}
\label{sec:LL}

In order to describe the low-temperature and low-energy properties of the system, we follow the standard procedure of linearizing the fermionic operators around the Fermi surface and bosonizing the result \cite{Giamarchi}. First we write the fermionic operators as:
\begin{align}
\psi(x) &= \sum_{r = \pm 1} e^{i r k_F x} \psi_r(x) \nonumber \\
\psi_r(x)  &= \frac{1}{\sqrt{L}}\sum_{k} \psi(r k_F + k)e^{i k x} 
\end{align}
where $k_F$ is the Fermi wavenumber and $\mu = k_F^2/2m$. In terms of these chiral modes the Hamiltonian in Eq.~(\ref{eq:H_xspace}) is written as:
\begin{align}
H_0 &= \sum_r \int dx \ \psi_r^{\dagger}(x)\left[-i r v_F \partial_x - \frac{\partial_x^2}{2m}\right] \psi_r(x) \label{eq:H0} \\
V_0 &= \int dx  g k_F^2 n_R(x) n_L(x) \label{eq:V0}\\
V_1 &= \sum_r \int dx \ i r g k_F \left[\psi^{\dagger}_r(x) \overleftrightarrow{\partial_x} \psi_r(x) n_{\bar{r}}(x) \right] \label{eq:V1} \\
V_2 &= \int d x \ g  \psi_R^{\dagger}(x) \overleftrightarrow{\partial_x} \psi_L^{\dagger}(x) \psi_L(x) \overleftrightarrow{\partial_x}\psi_R(x) \label{eq:V2} \\
V_2' &= \sum_r\int d x \ g  \psi_r^{\dagger}(x) \overleftrightarrow{\partial_x} \psi_r^{\dagger}(x) \psi_r(x) \overleftrightarrow{\partial_x}\psi_r(x) \label{eq:V2prime}
\end{align}
In Eqs.~(\ref{eq:H0}-\ref{eq:V2prime}) $n_r(x) = \psi_r^{\dagger}(x) \psi_r(x)$ is the density of $r$-moving fermions. We have also neglected writing the implicit normal ordering operator for clarity.

Secondly we write the chiral fermionic operators, $\psi_r(x)$ as:
\begin{align}
\psi_r(x) &= \frac{1}{\sqrt{2\pi \alpha}}e^{\sqrt{2\pi}i \phi_r(x)} \nonumber \\
&= \frac{1}{\sqrt{L}} e^{-\sqrt{2\pi} i \phi_r^{+}(x) }e^{-\sqrt{2\pi} i \phi_r^{-}(x) }
\label{eq:psi}
\end{align}
where $\phi_r(x)$ is the chiral bosonic operator, $\alpha$ is a short distance cut-off that we take to zero at the end of the calculation, and $L$ is the length of the system. We have also neglected the Klein factors for simplicity. In the second equality we write the bosonic field in terms of their annihilation, $\phi^-_r(x)$, and creation, $\phi^+_r(x)$ parts.  The commutation relation between the creation and annihilation operators is:
\begin{equation}
\left[\phi_r^{-}(x),\phi_{r'}^{+}(y)\right] = -\frac{1}{2\pi} \ln \left[\frac{-2\pi i r}{L} (x-y+i\alpha)\right]\delta_{r,r'}.
\label{eq:commutation}
\end{equation}

From Eqs.~(\ref{eq:psi}-\ref{eq:commutation}), one can then write Eqs.~(\ref{eq:H0}-\ref{eq:V2prime}) in terms of the bosonic field $\phi_r(x)$ by the method of point-splitting. The result is:
\begin{align}
H &\approx  H_{LL} + H_{bc} + V_{1} \nonumber \\
H_{LL}&= \int dx \left\lbrace \frac{v_F}{2} \left[\left(\partial_x \phi_R\right)^2 + \left(\partial_x \phi_L\right)^2\right] \right. \nonumber \\
& \left. - \frac{gk_F^2}{2\pi} \left(\partial_x \phi_R\right)\left(\partial_x \phi_L\right)\right\rbrace \label{eq:HLL_app} \\
H_{bc} &=\frac{\sqrt{2\pi}}{6m} \int dx \left[\left(\partial_x \phi_L\right)^3 - \left(\partial_x \phi_R\right)^3\right] \label{eq:Hbc_LL} \\
V_{1} &= \frac{g k_F}{2\sqrt{2\pi}} \int dx \left[ \left(\partial_x \phi_R \right)^2 \left(\partial_x \phi_L\right)-L \leftrightarrow R\right] \label{eq:V1_LL}
\end{align}

As we will discuss below Eq.~(\ref{eq:HLL_app}) gives the standard Luttinger Liquid model that describes sound waves with linear dispersions. The contributions from Eqs.~(\ref{eq:Hbc_LL}-\ref{eq:V1_LL}) give the leading irrelevant interactions.

In defining $H$ we have neglected the contributions due to $V_2$ and $V_2'$. These terms are more irrelevant than $H_{bc}$ and $V_1$. To facilitate this discussion consider the bosonized form of $V_2$:
\begin{align}
V_2 &= -\frac{g}{24\pi^2} \int dx  \left[\partial_x^3 \phi_R \partial_x \phi_L + \phi_x \phi_R \partial_x^3 \phi_L\right] \nonumber \\
& + \frac{g}{4\pi} \int dx \left(\partial_x^2 \phi_R \right)\left(\partial_x^2 \phi_L \right) \nonumber \\
&+ \frac{g}{4\pi} \int dx \left(\partial_x \phi_R \right)^2 \left(\partial_x \phi_L \right)^2 \nonumber  \\
&- \frac{g}{6\pi} \int dx \left[\left(\partial_x \phi_R\right)^3 \partial_x \phi_L + \partial_x \phi_R \left(\partial_x \phi_L\right)^3 \right] \label{eq:V2_LL}
\end{align}
$V_2$ has two distinct effects. First, $V_2$ modifies the dispersion of the sound waves. Equivalently one can say that the LL parameter and the renormalized sound velocity, defined below, become momentum dependent. As we will discuss later, this momentum dependence is not important to the discussion of the bulk  viscosity and can be neglected. Secondly $V_2$ generates interactions that are quartic in the bosonic fields. These interactions are more irrelevant than the cubic interactions and can also be ignored.  

To bring Eqs.~(\ref{eq:HLL_app}-\ref{eq:V1_LL}) to the standard LL form define:
\begin{align}
\phi_L &= \frac{\tilde{\theta} + \tilde{\phi}}{\sqrt{2\pi}} & \phi_R &= \frac{\tilde{\theta} - \tilde{\phi}}{\sqrt{2\pi}} 
\end{align}
In terms of these new fields one obtains:
\begin{align}
H_{LL} &= \int dx \ \frac{v}{2\pi} \left[\mathcal{K} \left(\partial_x \tilde{\theta}\right)^2 + \frac{1}{\mathcal{K}} \left(\partial_x \tilde{\phi}\right)^2\right] \nonumber \\
H_{bc} &= \int dx \ \frac{1}{6\pi m } \left[ 3 \left(\partial_x \tilde{\theta}\right)^2 \left(\partial_x \tilde{\phi}\right) +  \left(\partial_x \tilde{\phi}\right)^3\right] \nonumber \\
V_1 &= \int dx \ \frac{gk_F}{(2\pi)^2} \left[ -\left(\partial_x \tilde{\theta}\right)^2 \left(\partial_x \tilde{\phi}\right) + \left(\partial_x \tilde{\phi}\right)^3\right]
\end{align}
where the renormalized $v$ and LL paramete, $\mathcal{K}$ are defined as:
\begin{align}
v &= v_F \sqrt{1- \left(\frac{mg k_F}{2\pi}\right)^2} & \mathcal{K} &= \sqrt{\frac{1- m g k_F / 2\pi}{1+ m g k_F / 2\pi}}
\label{eq:renorm_LL}
\end{align}

For our purposes it is more convenient to define renormalized chiral modes via \cite{Pereira07}:
\begin{align}
\varphi_L(x) &= \sqrt{\frac{1}{2\pi}}\left(\mathcal{K}\tilde{\theta}(x) + \frac{1}{\mathcal{K}}\tilde{\phi}(x)\right) \label{eq:dressed_chiralL}\\\varphi_R(x) &= \sqrt{\frac{1}{2\pi}}\left(\mathcal{K}\tilde{\theta}(x) - \frac{1}{\mathcal{K}}\tilde{\phi}(x)\right)
\label{eq:dressed_chiralR}
\end{align}
In terms of the dressed chiral modes, Eq.~(\ref{eq:dressed_chiralL}-\ref{eq:dressed_chiralR}), the Hamiltonian becomes:
\begin{align}
H_{LL} &= \int dx \frac{v}{2} \left[\left(\partial_x \varphi_R\right)^2 + \left(\partial_x \varphi_L\right)^2\right] \label{eq:HLL_final} \\
V_- &= \eta_- \int dx \left((\partial_x\varphi_L)^3-(\partial_x\varphi_R)^3\right) \label{eq:V_-} \\
V_+ &= \eta_+  \int dx \left((\partial_x\varphi_L)^2 \partial_x\varphi_R - (\partial_x\varphi_R)^2 \partial_x\varphi_L \right) \label{eq:V_+} \\
\eta_- &= \frac{\sqrt{2\pi}\tilde{\eta}_-}{6} =  \sqrt{\frac{2\pi}{\mathcal{K}}} \frac{1}{24 m} \left[ 3 + \mathcal{K}^2 + \frac{3 mg k_F}{2\pi} \left(\mathcal{K}^2-1\right)\right] \label{eq:eta_-} \\
\eta_+ &=\frac{\sqrt{2\pi}\tilde{\eta}_+}{6} = \sqrt{\frac{2\pi}{\mathcal{K}}} \frac{1}{8  m}  \left[ (1- \mathcal{K}^2) - \frac{mg k_F}{2\pi} \left(1+3\mathcal{K}^2\right)\right] \label{eq:eta_+}
\end{align}

In this representation the density fluctuations are given by:
\begin{equation}
n(x) = - \sqrt{\frac{\mathcal{K}}{2\pi}}\sum_{r = \pm 1} r \partial_x \varphi_r(x) 
\end{equation}
Similarly, we define the current and stress-tensor according to:
\begin{align}
0&=\partial_t n(x,t) + \partial_x J(x,t) \nonumber \\
0&= \partial_t J(x,t) + \partial_x \Pi(x,t)
\label{eq:continuity}
\end{align}
From these definitions, and the Heisenberg equations of motion, the current $J(x)$ and stress-tensor $\Pi(x)$ operators are found to be:
\begin{align}
J(x) &= \sum_{r} -\sqrt{\frac{\mathcal{K}}{2\pi}} \left[ v \partial_x \varphi_r(x) \right. \nonumber \\
&\left.- \frac{\sqrt{2\pi}r}{6} \left(3\tilde{\eta}_- + \tilde{\eta}_+\right) \left(\partial_x \varphi_r(x)\right)^2\right] \label{eq:J(x)} \\
\Pi(x) &= \sum_r -\sqrt{\frac{\mathcal{K}}{2\pi}} \left[ v^2 r \partial_x \varphi_r(x) \right.\nonumber \\
&\left. - 2v \frac{\sqrt{2\pi} \tilde{\eta}_+}{6} \left(\partial_x \varphi_r(x)\partial_x \varphi_{\bar{r}} (x) - \left(\partial_x \varphi_r (x)\right)^2\right)  \right. \nonumber \\
&\left. - \frac{4\sqrt{2\pi}}{6}(3\tilde{\eta}_- + \tilde{\eta}_+) \mathcal{H}(x)\right] \label{eq:Pi(x)}
\end{align}

The current form of $\Pi(x)$ can not be transparently connected to Eq.~(\ref{def:pi}). That said,  when one performs the trace in the non-interacting limit the stress-tensor satisfies:
\begin{equation}
\lim_{\ell \to 0} \Pi = \lim_{\ell \to 0} \int dx \Pi(x) = \frac{2H}{m}.
\end{equation}
This is the relation expected by conformal symmetry, up to a factor of $m$.

\section{Calculation of the Bulk Viscosity via Stress-Tensor}

Next we calculate the bulk viscosity Eq.~(\ref{eq:zeta_def_final}) in both the low- and high-temperature limits. At low-temperatures, the physics is described by the LL model presented in Eqs.~(\ref{eq:HLL_final}-\ref{eq:eta_+}). In order to calculate the bulk viscosity, we consider the trace of the stress-tensor:
\begin{align}
\Pi(t) = \sqrt{\frac{\mathcal{K}}{2\pi}} &\left[\frac{4\sqrt{2\pi}}{6} \left(3\tilde{\eta}_- + \tilde{\eta}_+\right) H \right.\nonumber \\
&\left. - \frac{2v\sqrt{2\pi}\tilde{\eta}_+}{6} \sum_r \int dx \left( \partial_x \varphi_r(x,t)\partial_x \varphi_{\bar{r}}(x,t) \right. \right.  \nonumber \\
&\left. \left. - \left(\partial_x \varphi_r(x,t)\right)^2 \right)\right]
\label{eq:PI}
\end{align}

Since the commutator between the Hamiltonian and any operator vanishes in thermal equilibrium, one only needs to worry about the second line of Eq.~(\ref{eq:PI}) in  calculating Eq.~(\ref{eq:zeta_def_final}). To explicitly calculate Eq.~(\ref{eq:zeta_def_final}) we use finite temperature field theory. In doing so we replace the retarded correlation function in Eq.~(\ref{eq:PiPi_retarded}) with the imaginary time-ordered Green's function and evaluate it to leading order in perturbation theory. The result is:
\begin{align}
\chi_{\Pi,\Pi}(q, i \omega_n) &= \mathcal{K} \frac{v^2\tilde{\eta}_+^2}{9}\sum_r \frac{1}{\beta}\sum_{i \eta_n} \int_{-\infty}^{\infty} \frac{dp}{2\pi} \nonumber \\
& \left[D_r(p, i \eta_n) D_{\bar{r}}(-p, -i \eta_n+ i \omega_n) \right. \nonumber \\
&\left.+3 D_r(p, i \eta_n) D_r(-p, -i \eta_n + i \omega_n)\right]
\label{eq:PiPi_Im_Time}
\end{align}
where the bare bosonic propagator, $D_r(q, i\omega_n)$ is given by:
\begin{equation}
D_r(q,i\omega_n) = \frac{-rq}{i\omega_n -r v q}
\end{equation}

After performing the frequency summation in Eq.~(\ref{eq:PiPi_Im_Time}) only one term survives:
\begin{align}
\chi_{\Pi,\Pi}(q, i \omega_n) &= -\mathcal{K} \frac{v^2\tilde{\eta}_+^2}{9}\sum_r  \nonumber \\
& \int_{-\infty}^{\infty} \frac{dp}{2\pi}\frac{p^2}{i \omega_n - 2 r v p} \left(1 + 2 n_B(r v p)\right)
\label{eq:LL_chi}
\end{align}
where $n_b(x) = \left[\exp(\beta x)-1\right]^{-1}$ is the Bose-Einstein distribution.

The bulk viscosity can now be evaluated by substituting the result of Eq.~(\ref{eq:LL_chi}) into Eq.~(\ref{eq:zeta_def_final}) and then performing the analytic continuation, $i \omega_n \to \omega + i \delta$. This procedure gives the following expression for the bulk viscosity:
\begin{equation}
\zeta(\omega)  \approx \mathcal{K} \frac{\tilde{\eta}_+^2}{72v} \frac{\omega}{\tanh\left(\frac{\beta \omega}{4}\right)}
\end{equation}

In the weakly interacting limit this gives:
\begin{equation}
\zeta(\omega) \approx \frac{m \ell^2 k_F}{8\pi^2} \frac{\omega}{\tanh\left(\frac{\omega}{4 T}\right)}
\label{eq:zeta_final}
\end{equation}
where we note that the definition of $\Pi$ used in this work, Eq.~(\ref{eq:Pi(x)}) differs from the traditional definition by a factor of $m$. Thus we have added an additional factor of $m^2$, which is correct to leading order in $k_F \ell$.

\section{Calculation of the Bulk Viscosity Using Microscopic Theory}

In this section we present a derivation of the bulk viscosity in the high temperature limit by using the full microscopic theory.  We begin with Eq.~(\ref{eq:zeta_def_final}). Although the bulk viscosity is defined in terms of the stress-tensor correlation function, we note that the commutator of the Hamiltonian with any operator vanishes in thermal Equilibrium. Following Eq.~(\ref{def:pi}), the bulk viscosity can also be defined in terms of the the contact correlation function \cite{Nishida19, Maki20b, Enss20, Hofmann20}:
\begin{equation}
\zeta(\omega) =  \frac{Im \left[\chi_{C,C}(\omega)-\chi_{C,C}(0)\right]}{\ell^2\omega}
\label{eq:CC_zeta}
\end{equation}
where:
\begin{equation}
\chi_{C,C}(\omega)  = \int_{-\infty}^{\infty} dt e^{i (\omega+i \delta) t} i \theta(t) \langle \left[C_{\ell}(t), C_{\ell}(0)\right] \rangle
\label{eq:CC_retarded}
\end{equation}
and the contact operator $C_{\ell}$ is given by Eq.~(\ref{def:contact}).

The evaluation of Eq.~(\ref{eq:CC_retarded}) can be calculated using finite temperature field theory. According to the virial expansion, the leading contribution to Eq.~(\ref{eq:CC_zeta}) comes from the pair-pair propagator:
\begin{align}
\zeta(\omega) \approx \int_{-\infty}^{\infty} &\frac{dQ}{2\pi} \int_{-\infty}^{\infty}\frac{dx}{\pi} \frac{n_B(x)-n_B(x+\omega)}{\ell^2\omega}\nonumber \\
&Im \left[\frac{T(Q,x-i \delta)}{4}\right]Im \left[\frac{T(Q,x+\omega-i \delta)}{4}\right]
\label{eq:zeta_micro}
\end{align}
with $n_B(x)$ is the Bose-Einstein distribution at temperature $T$, and $T(Q,z)$ is the T-matrix, Eq.~(\ref{eq:T-matrix}). 

In the high temperature limit the chemical potential is large and negative, and an expansion in terms of the fugacity, $z = e^{\beta \mu}$, is permissible. In the high-temperature limit one finds:
\begin{align}
\zeta(\omega) &\approx z^2 \frac{1-e^{-\beta \omega}}{\ell^2\omega}\int_{-\infty}^{\infty} \frac{dQ}{2\pi} \int_{-\infty}^{\infty} \frac{dx}{\pi} e^{-\beta \frac{Q^2}{4} -\beta x}\nonumber \\
&Im \left[\frac{T(x-i \delta)}{4}\right]Im \left[\frac{T(x+\omega-i \delta)}{4}\right]
\label{eq:zeta_highT}
\end{align}
and
\begin{align}
&\left(\frac{T(x-i\delta)}{4}\right)^{-1} = \frac{1}{\ell}  + \sqrt{-x + i \delta}
\end{align}

In the weakly interacting limit one finds:
\begin{equation}
\zeta(\omega \to 0) \approx \frac{2}{\pi^{5/2}} \left( k_F \ell\right)^2 T^{1/2}
\label{eq:zeta_highT_final}
\end{equation}
To obtain Eq.~(\ref{eq:zeta_highT_final}) we have used the equation of state to relate the fugacity to the density, $n$, and the Fermi momentum, $k_F$:
\begin{align}
z &= n \sqrt{2\pi \beta} & n &= \frac{k_F}{\pi}
\end{align}

Conversely, in the resonantly interacting limit, $\ell^{-1} =0$, one obtains:
\begin{align}
\lim_{\ell^{-1} \to 0} \lim_{\omega \to 0}\zeta(\omega) &\approx 2\frac{k_F^2}{\pi^{5/2}} \frac{1}{\ell^2}\frac{1}{T^{3/2}} \ln \left(m \ell^2 T\right) \\
\lim_{\omega \to 0} \lim_{\ell^{-1} \to 0}\zeta(\omega) &\approx  2\frac{k_F^2}{\pi^{5/2}} \frac{1}{\ell^2}\frac{1}{T^{3/2}} \ln \left(\frac{4T}{\omega}\right)
\label{eq:zeta_highT_final2}
\end{align}

We briefly note that Eq.~(\ref{eq:zeta_micro}) also describes the low-temperature bulk viscosity to leading order in perturbation theory. A direct evaluation of  Eq.~(\ref{eq:zeta_micro}) using the microscopic model at low-temperatures gives results consistent with Eq.~(\ref{eq:zeta_final}) in the zero-frequency limit. However, one needs to use the two-body T-matrix in the presence of the many-body background. We won't provide the details here, but provide a similar analysis for the Lieb-Liniger model below. This further confirms that the LL model provides an accurate evaluation of the bulk viscosity.

\section{Calculation of the Scattering rate in Kinetic Theory}

In this section we consider using a pseudo kinetic theory to evaluate the bulk viscosity. In this approach we examine the response of Fermionic quasiparticles to a two-body dephasing process beyond the LL model. Although these processes do not physically relax the momentum, we find results consistent with the more microscopic definition of the bulk viscosity. This approach is in contrast to previous studies where the bulk viscosity is explicitly related to three-body processes which do relax the momentum of the system  \cite{Matveev17, DeGottardi20}, but provide answers inconsistent with conformal symmetry \cite{Son07}.

Consider the following Hamiltonian for fermionic quasiparticles:
\begin{align}
H &= H_0 + V_+ \nonumber \\
H_0 &= \sum_r\int dx \colon \psi^{\dagger}_r(x)\left(-i r v \partial_x\right) \psi_r(x) +\frac{1}{\tilde{m}} \partial_x \psi^{\dagger}(x) \partial_x \psi(x) \colon  \nonumber \\
V_+  &= \int dx \sum_r \colon i r\eta_+ \psi^{\dagger}_r(x) \frac{\overleftarrow{\partial_x}-\overrightarrow{\partial_x}}{2} \psi_r(x) \psi_{\bar{r}}^{\dagger}(x) \psi_{\bar{r}}(x) \colon
\label{eq:boltz}
\end{align}
In Eq.~(\ref{eq:boltz}) we have already included the LL renormalization effects and the non-perturbative summation of the $V_-$ interactions in order to define quasiparticles with  velocity, $v$, Eq.~(\ref{eq:renorm_LL}), and effective mass $\tilde{m}^{-1} = \tilde{\eta}_-$, Eq.~(\ref{eq:eta_-}). The interactions beyond the LL model couple right and left moving fermions together with an interaction strength $\eta_+$ from Eq.~(\ref{eq:eta_+}).

To examine the bulk viscosity we subject the system to a slowly changing velocity field $u$. Because of the interactions, there will necessarily be power dissipated. In the limit of small velocity gradients, one expects the power dissipated to be of the form:
\begin{equation}
W = \zeta L (\partial_x u)^2
\end{equation}
where $\zeta$ is the zero-frequency limit of the bulk viscosity spectral function. 

From standard thermodynamic arguments the power dissipated is $W = T \dot{S}$, where $S$ is the entropy. The entropy can be explicitly calculated using the fermionic statistics of the quasiparticles:
\begin{equation}
S = - \sum_{r,p} \left[n_{r,p} \ln(n_{r,p}) + (1-n_{r,p}) \ln(1-n_{r,p}) \right]
\end{equation}
where $n_{r,p}$ is the time-dependent distribution function of a $r$-moving quasiparticle with momentum $p$. Assuming the velocity gradient is weak enough, the momentum distribution can be expanded as $n_{r,p} = n_{r,p}^0 + \delta n_{r,p}$, with $\delta n_{r,p} \propto \partial_x u$. The equilibrium distribution function can be written as:
\begin{equation}
n_{r,p}^0 = \left(\exp\left[\beta\left(r v p + \frac{p^2}{2\tilde{m}} -\Delta\mu\right) \right]+1\right)^{-1}.
\label{eq_n}
\end{equation}
In defining Eq.~(\ref{eq_n}), we note that $\Delta\mu$ is the shift in the chemical potential away from its zero-temperature value: $\Delta \mu = \mu(T) -\mu(T=0)$, as the zero-temperature chemical potential is used in writing down the linearized Hamiltonian. From the Sommerfeld expansion one finds:
\begin{equation}
\Delta\mu = \frac{\pi^2 T^2}{12 \tilde{m} v^2}.
\end{equation}
Then expanding $W$ to linear order in $\delta n_{r,p}$ gives:
\begin{equation}
W = - T \sum_{r,p} \frac{ \dot{n}_{r,p}^0 \delta n_{r,p}}{n_{r,p}^0 (1-n_{r,p}^0)}.
\label{eq:work}
\end{equation}
where $\dot{n}_{r,p}^0$ represents the total time derivative of $n_{r,p}^0$.

To proceed further we need to evaluate the time-dependence of the equilibrium distribution function: $\dot{n}_{r,p}^0$. Following Refs.~\cite{Matveev17, DeGottardi20} we note that the velocity gradient can be imagined as a box with a size $L(t)$ expanding such that: $\partial_x u = \dot{L}(t)/L(t)$. In this picture the time dependence of the density and momentum $p$ are trivial:
\begin{align}
\frac{\dot{n}}{n} &= - \partial_x u & \frac{\dot{p}}{p} &= - \partial_x u
\end{align}
In order to obtain the time-dependence of the temperature and the chemical potential, we note that the quasi-particle collisions preserve the quasi-particle number and energy:
\begin{align}
\sum_p \dot{n}_{r,p}(t) &= 0 & \sum_p \epsilon_r(p)\dot{n}_{r,p}(t) &= 0 
\label{eq:conservations}
\end{align}
where $\epsilon_{r}(p) \approx r v p$. Requiring the equilibrium distribution function satisfy Eq.~(\ref{eq:conservations}), one finds the following time-dependences for $\beta=1/T$ and $\Delta \mu$:
\begin{align}
\frac{\dot{\beta}}{\beta} &= \left[n \partial_n \ln(n v) \right] \partial_x u & \frac{\dot{\Delta\mu}}{\Delta\mu} &= \left[n \partial_n \ln\left(\frac{\tilde{m}}{n^2}\right) \right] \partial_x u 
\end{align}
which is valid to first order in $1/(\beta \tilde{m} v^2)$. Finally we can obtain the time-derivative of the distribution function as:
\begin{equation}
\dot{n}_{r,p}^0 = n_{r,p}^0(1-n_p^0) \frac{6v^2 p^2 - \pi^2 T^2}{12 \tilde{m}v^2 T} \left( n \partial_n \ln(\nu)\right) \partial_x u 
\label{eq:dot_n}
\end{equation}
where $\nu = \frac{\pi n}{\tilde{m}v}$.

Next we need to calculate $\delta n_{r,p}$. This is done using the relaxation time-approximation: $\delta n_{r,p} = -\tau_{r,p} \dot{n}^0_{r,p}$, where $\tau_{r,p}$ is the scattering time associated with the two-body interactions beyond the LL model, $V_+$. Using the relaxation time-approximation the bulk viscosity is found to be:
\begin{equation}
\zeta = \frac{T}{L} \sum_p \frac{(\dot{n}_p^0)^2 }{n_p^0(1-n_p^0)} \tau.
\label{eq:bulk_tau}
\end{equation}

The definition of $\tau$ follows from Fermi's golden rule. Consider the scattering time for a right moving fermion with momentum $p$.
\begin{equation}
\frac{1}{\tau} = \sum_{k_1,k_2,k_1',k_2'}\sum_{r'} \delta_{p, k_1} W_{1,2}^{1',2'} n_L(k_2) (1-n_{r'}(k_1'))(1-n_{\bar{r}'}(k_2'))
\label{eq:tau_p}
\end{equation}
where $\bar{r} = -r$ and
\begin{align}
W_{1,2}^{1',2'} = 2\pi &\delta\left(\epsilon_{r}(k_1) + \epsilon_{\bar{r}}(k_2) - \epsilon_{r'}(k_1') - \epsilon_{\bar{r}'}(k_2')\right) \nonumber \\
&| \langle k_1' \ r_1' , k_2' \ r_2' | V_+ | k_1 R, k_2 L\rangle |^2.
\end{align}

The matrix element due to $V_+$ is given by:
\begin{align}
\langle k_1' \ r_1' , k_2' \ r_2' | & V_+ | k_1 R, k_2 L\rangle = \nonumber \\
&\frac{\eta_+}{L}\left[(q+q')\delta_{R,r_1'} + (q-q') \delta_{R,\bar{r}_1'}\right].
\end{align}
A straightforward calculation provides:
\begin{align}
\frac{1}{\tau} = \frac{4\eta_+^2}{v}\int \frac{dq}{2\pi} q^2 n_L^0(p-2q)(1- n_L^0(p-2q)) (1-n_R^0(p)),
\label{eq:tau_p_eval}
\end{align}
In order to evaluate Eq.~(\ref{eq:tau_p_eval}) we need to perform the normal ordering, which essentially subtracts off the $T=0$ contribution. However, the $T=0$ contribution is identically zero. Therefore the relaxation time is finally given by:
\begin{equation}
\frac{1}{\tau} = \frac{\eta_+^2T^3}{3v^4}
\label{eq:tau_final}
\end{equation} 
Substituting Eqs.~(\ref{eq:dot_n}) and  Eq.~(\ref{eq:tau_final}) into Eq.~(\ref{eq:bulk_tau}) gives the final result for the bulk viscosity:
\begin{equation}
\zeta \approx \frac{1}{24\pi} \frac{T}{v\tilde{m}^2\eta_+^2} \left[ n \partial_n \ln(\nu)\right]^2.
\end{equation}
From the definition of $\nu = \pi n / \tilde{m} v$, one can show that:
\begin{align}
n \partial_n \ln(\nu) = - \frac{5}{4} \left(\frac{2k_F \ell}{\pi}\right)^2.
\end{align}
This gives us the final expression for the bulk viscosity:
\begin{align}
\zeta &\approx \frac{25}{6\pi^4} m k_F \ell^2 T
\label{eq:zeta_pseudo_kinetic}
\end{align}
Eq.~(\ref{eq:zeta_pseudo_kinetic}) has the same functional form as the LL calculation using the microscopic definition of the bulk viscosity.

The main point to note is that one has to go beyond the standard kinetic theory calculation to produce results consistent with the microscopic evaluation of the bulk viscosity.

\section{Bulk Viscosity for the Lieb-Liniger Model}

Let us now consider the Lieb-Liniger model of interacting bosons:
\begin{align}
H &= \int dx \ \frac{1}{2m}\partial_x\phi^{\dagger}(x) \partial_x \phi(x) \nonumber \\
&+ \int dx \frac{g}{4} \phi^{\dagger}(x)\phi^{\dagger}(x)\phi(x)\phi(x)
\end{align}
where $\phi(x)$ is the bosonic operator, $m$ is the atomic mass and $g = -4/a$ relates the interaction strength to the scattering length, $a$.

In the high temeprature limit, the two-body T-matrix has the form:
%\begin{align}
%&\left(\frac{T(Q,x-i \delta)}{4} \right)^{-1}=  \nonumber \\
%&-a- \int \frac{dk}{\pi} \frac{1+n_B\left(\xi_{\frac{Q}{2}+k}\right) +n_B\left(\xi_{\frac{Q}{2}-k}\right)}{x- \xi_{\frac{Q}{2}+k}-\xi_{\frac{Q}{2}-k}-i\delta}
%\end{align}
\begin{align}
\left(\frac{T(x-i \delta)}{4} \right)^{-1}=  -a + \frac{1}{\sqrt{-x+ i \delta}}
\end{align}
where $\xi_k = \frac{k^2}{2m}-\mu$, and  $n_B(x) = \left(e^{\beta x}-1\right)^{-1}$ is the Bose-Einstein distribution.

We can again define the stress-tensor according to:
\begin{equation}
\Pi = 2 H + a C_a 
\end{equation}
where the  contact operator is:
\begin{equation}
C_a = \frac{\partial H}{\partial a}= \int dx \ \frac{g}{16}  \phi^{\dagger}(x)\phi^{\dagger}(x)\phi(x)\phi(x).
\end{equation}

From Eq.~(\ref{eq:zeta_def_app}) the bulk viscosity for the Lieb-Liniger model is defined as:
\begin{equation}
\zeta(\omega) = \frac{\chi_{\Pi\Pi}(\omega)-\chi_{\Pi\Pi}(0)}{i\omega} + \frac{2a \langle C_a\rangle}{i \omega}
\label{eq:bulk_visc_bosons_def}
\end{equation}
The extra term in Ref.~(\ref{eq:bulk_visc_bosons_def}) arises from the behaviour of the contact operator under a scale transformation. Since $g=-4/a$, the contact operator does not simply transform according to its scaling dimension of $-3$ but obeys the modified relation:
\begin{equation}
\left. \frac{dC_a}{db}\right|_{b=0} = -C_a.
\end{equation}
This leads to the following behavior of $\Pi$ under a scale transformation:
\begin{equation}
\left. \frac{d\Pi_b}{db}\right|_{b=0} = -2 \Pi -a C_a.
\end{equation}
This ought to be compared to how the expectation value $\langle \Pi \rangle$ changes under a scale transformation:
\begin{equation}
\left.\frac{d }{db}\langle \Pi \rangle_b \right|_{b=0} =-2 \langle \Pi\rangle - 3 a \langle C_a \rangle - a^2 \frac{\partial \langle C_a \rangle}{\partial a}
\end{equation}

The last term can again be identified as $\chi_{\Pi\Pi}(0)$ via linear response theory, similar to  the Fermionic case. This substitution brings the original definition of the bulk viscosity, Eq.~(\ref{eq:zeta_def_app}), into the form of Eq.~(\ref{eq:bulk_visc_bosons_def}). The presence of the Drude peak is still consistent with scale symmetry, but suggests a divergent bulk viscosity spectral function at zero-frequency for fixed arbitrary interaction strengths. If one is considered with the real part of the bulk viscosity spectral function, the Drude peak only appears at zero-frequency. In this work, we always assume that the frequency is finite but vanishingly small, so as to ignore the Drude peak. 

At leading order the bulk viscosity can be written as:
\begin{align}
\zeta(\omega) \approx \frac{a^2}{\pi \beta} & \int_{-\infty}^{\infty}\frac{dx}{\pi} \frac{n_B(x)-n_B(x+\omega)}{\omega}\nonumber \\
&Im \left[\frac{T(x-i \delta)}{4}\right]Im \left[\frac{T(x+\omega-i \delta)}{4}\right]
\label{eq:zeta_micro_bosons}
\end{align}
where $T(x-i\delta)$ is the two-body T-matrix defined in the vacuum. Eq.~(\ref{eq:zeta_micro_bosons}) is valid in the high-temperature limit for arbitrary interaction strengths. Near resonance $na \ll 1$ the bulk viscosity can be analytically evaluated:
\begin{equation}
\zeta(\omega \to 0) \approx \frac{2}{\pi^{5/2}} \left( k_F a\right)^2 T^{1/2}
\label{eq:zeta_highT_final_bosons}
\end{equation}
where we have used $n= k_F/\pi$, and similarly for the weakly interacting limit $na \gg 1$:
\begin{equation}
\lim_{\omega \to 0} \lim_{a^{-1} \to 0}\zeta(\omega) \approx  2\frac{k_F^2}{\pi^{5/2}} \frac{1}{a^2}\frac{1}{T^{3/2}} \ln \left(\frac{4T}{\omega}\right)
\label{eq:zeta_highT_final_bosons2}
\end{equation}

Upon inspection of Eqs.~(\ref{eq:zeta_highT_final_bosons}-\ref{eq:zeta_highT_final_bosons2}), one finds that they are equivalent to Eqs.~(\ref{eq:zeta_highT_final},\ref{eq:zeta_highT_final2}) in the opposing limit. This allows one to establish the duality relations:
\begin{equation}
\zeta_F(\ell) = \zeta_B \left(\frac{1}{a}\right).
\end{equation}
Hence the bulk viscosity for a strongly (weakly) interacting Fermi gas is equal to the bulk viscosity of a weakly (strongly) interacting Bose gas.

\begin{figure}
\includegraphics[scale = 0.6]{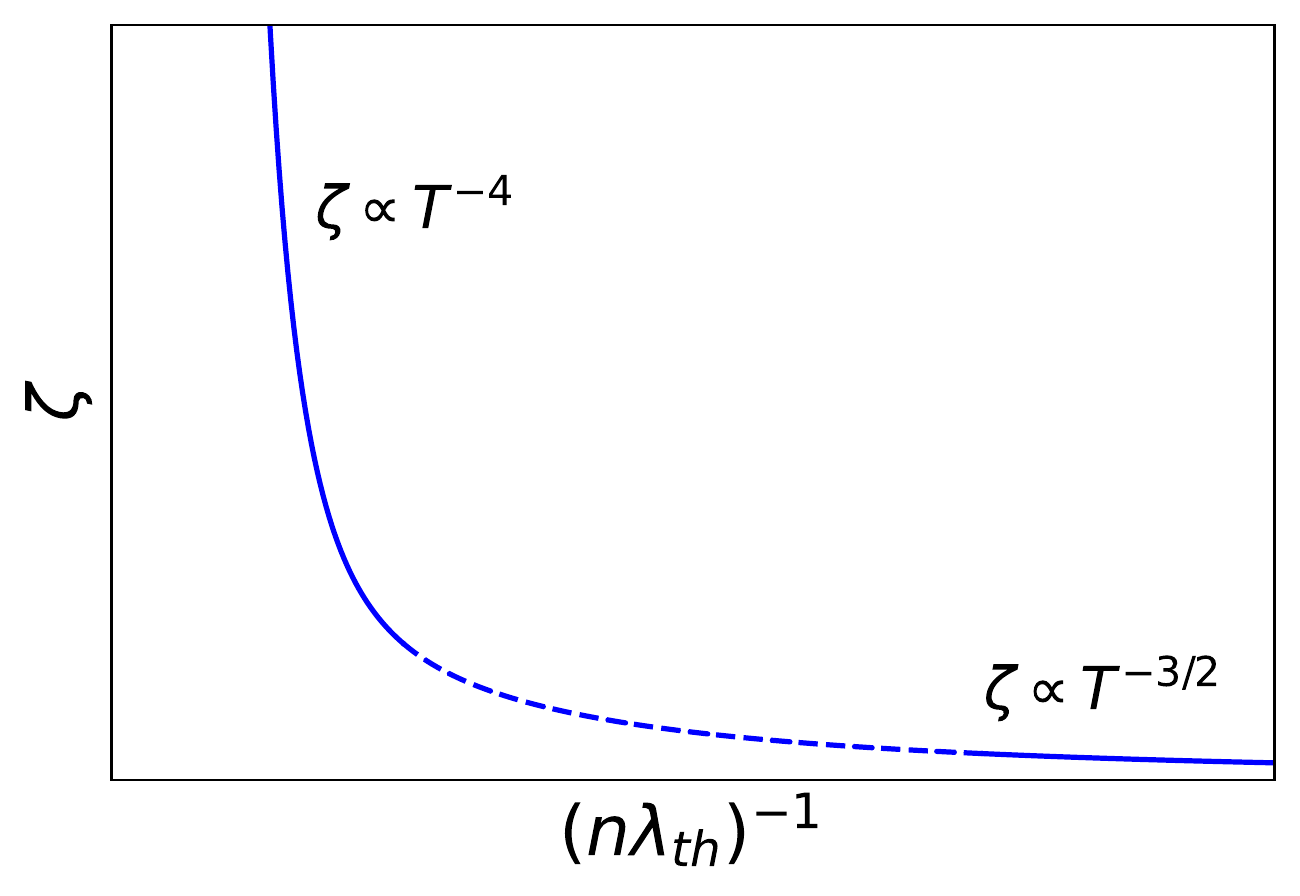}
\caption{Bulk viscosity of a spin-polarized Fermi gas near resonance. At high temperatures, Eq.~(\ref{eq:zeta_highT_final2}), the bulk viscosity vanishes like $\ell^{-2} n^2 T^{-3/2}$, while for low temperatures the bulk viscosity diverges as: $\ell^{-2} n^7 T^{-4}$. From the duality relations, this plot also describes the bulk viscosity for a weakly interacting 1D bose gas.}
\label{fig:res}
\end{figure}

In the low-temperature limit the bulk viscosity can be evaluated at low-temperatures. The only required modification to Eq.~(\ref{eq:zeta_micro_bosons}) is to use the T-matrix calculated within the many-body background. To leading order in perturbation theory one can write the many-body T-matrix as:
\begin{align}
Im\left[\frac{T(Q,x-i\delta)}{4}\right] &\approx \nonumber \\
\frac{1}{a^2} \int \frac{dk}{\pi} & \left[1+n_B\left(\xi_{\frac{Q}{2}+k}\right) +n_B\left(\xi_{\frac{Q}{2}-k}\right)\right] \nonumber \\
&\times \pi \delta(x- \xi_{\frac{Q}{2}+k}-\xi_{\frac{Q}{2}-k}-i\delta) 
\end{align}
From this expression, one can show that the integrand is dominated near $\epsilon \approx 0$ and $Q \approx 0$. Expanding near this point and evaluating the resulting integrals give:
\begin{equation}
\zeta_B(\omega\to 0) \approx \frac{16}{\pi^2}\frac{n^7}{a^2 T^4} \log\left(16\frac{T^2}{n^2\omega} \right)
\label{eq:weak_boson_zeta}
\end{equation}
In order to obtain Eq.~(\ref{eq:weak_boson_zeta}) we have used the non-interacting equation of state:
\begin{equation}
n = \frac{1}{\sqrt{2\pi \beta}} Li_{1/2}\left(e^{\beta \mu}\right)
\end{equation}
where $Li_s(x)$ is the polylogarithm function. As one approaches the zero-temperature limit for fixed density, one can show $\sqrt{2\beta}n = (-\beta \mu)^{-1/2}$, i.e. $\mu \to 0^-$ as $T \to 0$.

From Eqs.~(\ref{eq:zeta_highT_final2}) and (\ref{eq:weak_boson_zeta}) one can construct the bulk viscosity for arbitrary temperatures near resonance, as shown in Fig.~(\ref{fig:res}). From the duality relations Figs.~(\ref{fig:zeta_1}) and (\ref{fig:res})  describe the bulk viscosity for a 1D Bose gas near resonance and for weak interactions respectively.


\begin{thebibliography}{References}
\bibitem{V_minus_note} The sound modes do acquire a finite lifetime due to the presence of the intra-band scattering. However, this lifetime is in no way associated with the bulk viscosity and the dissipation of macroscopic flow. This is in contrast to the prediction from generalized hydrodynamics \cite{Samokhin98,Punk06}.
\end{thebibliography}
\end{document}